\documentclass[journal]{IEEEtran}

\usepackage[cmex10]{amsmath}
\usepackage{amssymb}
\usepackage{amsthm}
\usepackage{bm}
\usepackage{graphicx}
\usepackage{tikz}
\usetikzlibrary{arrows.meta,positioning,fit,calc}
\usepackage[caption=false,font=footnotesize]{subfig}
\usepackage{booktabs}
\usepackage{tabularx}
\usepackage{algorithm}
\usepackage{algorithmic}
\usepackage{cite}
\usepackage{stfloats}
\usepackage{placeins}
\usepackage{textcomp}
\usepackage{url}
\usepackage[colorlinks,citecolor=blue,linkcolor=blue,urlcolor=blue]{hyperref}

\newtheorem{proposition}{Proposition}

\newcommand{\vc}[1]{\bm{#1}}
\newcommand{\C}{\mathbb{C}}
\newcommand{\R}{\mathbb{R}}
\newcommand{\Hc}{\mathrm{H}}
\newcommand{\diag}{\operatorname{diag}}
\newcommand{\RePart}{\operatorname{Re}}

\definecolor{learnblue}{HTML}{176B87}
\definecolor{modelorange}{HTML}{C45A28}
\definecolor{sharedgray}{HTML}{4F5962}

\begin{document}

\title{Learning or Modeling? Virtual XL-RIS Observation Reconstruction for Near-Field Sensing}

\author{Tuo Wu, Jie Tang, Jiacheng Wang, Kangda Zhi, Maged Elkashlan, Matthew C. Valenti, \emph{Fellow, IEEE},\\ M\'erouane~Debbah, \emph{Fellow, IEEE}, and Kenneth Tong, \emph{Fellow, IEEE}
\thanks{Corresponding author: Jie Tang.}
\thanks{T. Wu and J. Tang are with the School of Electronic and Information Engineering,
South China University of Technology, Guangzhou 510640, China (E-mail: $\rm
wutuo@scut.edu.cn; eejtang@scut.edu.cn$). J. Wang is with the College of
Computing and Data Science, Nanyang Technological University, Singapore
639798 (E-mail: $\rm jiacheng.wang@ntu.edu.sg$). K. Zhi is with the School of Electrical Engineering and
Computer Science, Technical University of Berlin, 10623 Berlin, Germany
(E-mail: $\rm  k.zhi@tu-berlin.de$). M. Elkashlan is with the School of Electronic
Engineering and Computer Science, Queen Mary University of London, London
E1 4NS, U.K. (E-mail: $\rm maged.elkashlan@qmul.ac.uk$). M. C. Valenti is with the Lane Department of Computer Science and Electrical Engineering, West Virginia University, Morgantown, USA (E-mail: $\rm valenti@ieee.org$). M. Debbah is with  the Research Institute for Digital Future, Khalifa University, 127788 Abu Dhabi, UAE (E-mail: $\rm merouane.debbah@ku.ac.ae$). K. Tong is with the Department of Electronic and Electrical Engineering, University College London, London WC1E 7JE, U.K. (E-mail: $\rm k.tong@ucl.ac.uk$). }}

\markboth{}{}
\maketitle

\begin{abstract}
A passive extremely large-scale reconfigurable intelligent surface (XL-RIS)
offers a position-sensitive aperture for near-field sensing, but its
element-wise signals are not directly observed. Recovering this virtual RIS
observation (VOR) from fewer base-station measurements requires prior
information. This paper investigates when calibrated propagation geometry
or paired training data provides the more effective reconstruction prior.
For a common sensing operator and target signal, we develop off-grid
near-field sparse parametric reconstruction (OG-NF-SPR) and compare it with
compact and DenseNet-based deep reconstructors. OG-NF-SPR uses reflected-path
projection for position initialization, followed by Bayesian estimation of
all path gains and continuous position refinement. A causal extension
incorporates temporal correlation without trajectory labels. Controlled
simulations use near-field UAV tracking as an illustration and separately
examine noise, position-distribution shift, calibration error, and propagation
mismatch. Under matched propagation at 20 dB, OG-NF-SPR achieves approximately
67\% lower reconstruction NMSE than the two nominal deep networks.
At low SNR, statistical averaging is more effective than selecting a single
geometric explanation. Under sufficiently strong repeatable mismatch, a
same-size network trained on the mismatch family outperforms the nominal
physical estimator, whereas nominal training does not provide this benefit.
Temporal regularization also reduces position RMSE from 0.653 m to 0.225 m
on a shared trajectory. The results relate reconstruction performance to
measurement quality, calibration accuracy, and training coverage, providing
guidance on when geometry can replace supervised labels and when additional
data are useful.
\end{abstract}


\IEEEpeerreviewmaketitle

\newpage
\section{Introduction}
\IEEEPARstart{E}{xtremely} large-scale reconfigurable intelligent surfaces
(XL-RISs) provide an attractive way to introduce a large spatial aperture into
wireless sensing systems while keeping the surface passive. In the near
field, the phase curvature and amplitude variation across this aperture
depend on both source direction and range \cite{Guerra,HenkWymeersch1}.
These properties are useful for localization and tracking, including UAV
applications \cite{UAV3,UAV4}. However, the absence of element-wise receivers
creates an acquisition bottleneck: the aperture signal is encoded in a
smaller set of BS measurements through the RIS-to-BS channel and phase
profiles. Exploiting the large aperture therefore requires more than
increasing its number of elements; it requires recovering a useful spatial
representation from the measurements that are actually available.

RIS-assisted localization has been studied through information bounds,
receiving architectures, and phase-profile design
\cite{Elzan1,JiguangHe2,Mingan}. Near-field methods exploit power
fingerprints, hybrid surfaces, spherical waves, and practical element
responses \cite{Shubo,Xing,Chongwen1,Cuneyd,Reza}; tracking methods further
address spatial-wideband effects, precoding, and position-velocity estimation
\cite{Keykh1,Palmucci,NF2,NF3}. With larger apertures, spherical spreading and
phase curvature become increasingly relevant
\cite{WangXL2024,LuNFTut,ZhiJSAC2024,WangNFISAC}. These studies motivate
aperture-domain processing, but a passive surface still requires a means of
obtaining its spatial signal representation from BS measurements.

The \emph{virtual RIS observation} (VOR) provides such a representation. It is
the complex signal that would be observed across the RIS if its elements
could sample the incident field. Once reconstructed, it allows aperture-domain
feature extraction without adding receivers to the passive surface.
Earlier work examined the localization value of high-dimensional RIS
information \cite{Wu6}. Our previous studies developed reconstruction-assisted
localization \cite{Wu5} and a DenseNet-based VOR framework for near-field UAV
tracking \cite{WuJSTSP2026}. These studies motivate the use of VOR as an
intermediate representation, but its reconstruction introduces a distinct
design choice: should the BS rely on labelled examples of the deployment
or on an explicit propagation model? This choice determines whether the
system depends primarily on representative training data or accurate
geometric calibration, in addition to the online measurements.

VOR reconstruction is generally nonunique: with $Q<N$ measurements,
different $N$-element aperture signals can produce the same noiseless BS
observation. A prior must restrict the admissible signals. For example, the
samples of a spherical wave are jointly determined by a source position and
a complex gain, rather than $N$ independent unknowns. Physical estimation
uses this relation explicitly; learning captures the deployment's signal
distribution from paired BS/RIS samples.

Deep learning has been applied to channel estimation, detection, and angular
recovery \cite{ye2017power,hu2020deep,yang2019deep,huang2018deep}, as well as
fingerprint-based localization \cite{Bhattacherjee}. For VOR, squared-error
training ideally learns the conditional mean of the RIS target given the BS
input. Averaging plausible channels can reduce error when noise prevents a
reliable choice between source locations. The trained network also avoids
online geometric search. These benefits depend on representative labels:
accurate regression on the training distribution need not extend to new
positions or changed propagation conditions.

Physical modeling offers a different tradeoff. Near-field sparse estimation
uses angle--range dictionaries and spherical-wave structure
\cite{CuiNearFieldCE,ZhangSparseNF}; structured codebooks reduce beam-search
overhead \cite{ZhangLWC2022}, and continuous refinement mitigates grid error
\cite{MamandipoorNOMP}. In a calibrated XL-RIS deployment, a moving
line-of-sight (LoS) source and a few reflected paths similarly describe a
restricted family of aperture signals. Fitting this family requires no
supervised BS/RIS pairs and can evaluate locations absent from a training
set. However, reflected paths can obscure the source-dependent component
after compression, and a finite search grid limits position resolution.
Incorrect reflector coordinates or omitted propagation components introduce
a further problem: the fitted family may no longer describe the true signal.

These complementary strengths also motivate model-driven deep learning
\cite{HeModelDriven,ZhangModelDrivenNF}. Before combining the approaches,
however, one must distinguish the value of training coverage from that of
calibration. We compare both routes using the same sensing operator,
BS samples, and RIS target, while retaining their different prerequisites:
paired data for learning, and geometry and path statistics for modeling.
This controls the measurements, not the cost of obtaining the priors.
The networks learn RIS targets rather than physical-estimator outputs;
downstream tracking is evaluated separately.

We further distinguish changes that are often grouped together as
\emph{mismatch}. Moving a source outside the training distribution changes
coverage but need not change the propagation law. Supplying incorrect
reflector coordinates changes an estimator input without changing the
received signal. Adding unmodeled paths changes the physical signal itself.
The first test examines position generalization, the second calibration
sensitivity, and the third propagation mismatch. This distinction is essential
for interpreting a crossover between methods: a network that never uses
reflector coordinates is insensitive to errors in those coordinates, but that
alone does not show robustness to a changed channel.

To address this design question, we develop off-grid near-field sparse
parametric reconstruction (OG-NF-SPR) and study it alongside compact and
DenseNet-based deep reconstructors. The physical method separates two roles
of multipath: reflected components are removed to initialize the source
position, but all paths are retained when reconstructing the aperture
signal. This provides a label-free alternative whose dependence on
calibration can be tested directly. The main contributions are as follows.
\begin{itemize}
\item \emph{Reconstruction with explicit information requirements:} We place
physical fitting and learned inversion in a common compressed-observation
model. By separating the measured BS signal, the latent RIS signal, and the
source position, the formulation identifies what each estimator must infer
and which offline information it requires.
\item \emph{Off-grid full-aperture recovery:} We develop a reconstruction
procedure combining multipath-projected initialization, closed-form Bayesian
gain estimation, and safeguarded position refinement. All instantaneous path
gains remain unknown and are recovered from the current observation. A
causal extension exploits previous position estimates. We establish descent
of accepted objective values, give a local identifiability condition, and
characterize computational complexity.
\item \emph{Controlled comparisons and design implications:} We separately
vary measurement noise, position coverage, reflector metadata, and physical
propagation. Grid and off-grid estimators isolate refinement, while nominal
and domain-randomized networks of the same size isolate training coverage.
The results explain when calibrated geometry can substitute for supervised
labels and when representative data help compensate for a deficient nominal
model. Shared-trajectory experiments distinguish these reconstruction gains
from the contribution of temporal processing.
\end{itemize}

The remainder of this paper is organized as follows. Section~\ref{sec:system}
presents the system model and reconstruction problem.
Section~\ref{sec:algorithm} develops the reconstruction algorithms, and
Section~\ref{sec:analysis} provides theoretical analysis.
Section~\ref{sec:results} illustrates the methods through simulations of
near-field UAV tracking. Section~\ref{sec:conclusion} concludes the paper.

\section{System Model and Reconstruction Problem}\label{sec:system}

We consider a cooperative near-field sensing link in which a source
transmits a known pilot and the BS receives the signal through a passive
XL-RIS. Three quantities must be distinguished: the unknown source position
$\vc p$, the unobserved RIS signal $\vc y_{\mathrm r}$, and the measured BS
signal $\vc y$. Both reconstructors estimate $\vc y_{\mathrm r}$ from
$\vc y$. Position and path gains are intermediate unknowns for physical
fitting, whereas a learned reconstructor can predict the VOR directly.

\subsection{Virtual RIS Observation Model}

During one quasi-static snapshot, a source transmits a known narrowband pilot,
normalized to one without loss of generality. The \emph{virtual RIS
observation} $\vc y_{\mathrm r}\in\C^N$ is the vector that would be measured
across the $N$ RIS elements before passive phase control. The RIS cannot sample
this vector directly; it is the high-dimensional quantity to be reconstructed
from BS measurements. The XL-RIS center is located at
$\vc p_{\mathrm R}\in\R^3$, element $n$ is located at
$\vc p_n\in\R^3$, and the unknown source position is
$\vc p\in\mathcal P\subset\R^3$. For a candidate source location $\vc u$,
define
\begin{equation}
d_n(\vc u)=\|\vc u-\vc p_n\|_2,\quad
d_c(\vc u)=\|\vc u-\vc p_{\mathrm R}\|_2.
\label{eq:distances_paper}
\end{equation}
The center-referenced near-field atom is
\begin{equation}
[\vc a(\vc u)]_n=\frac{d_c(\vc u)}{d_n(\vc u)}
e^{-j\frac{2\pi}{\lambda}[d_n(\vc u)-d_c(\vc u)]}.
\label{eq:atom_paper}
\end{equation}
The amplitude ratio captures element-dependent spherical spreading. The common
phase and attenuation are absorbed into a complex path coefficient.

Suppose $S$ dominant single-bounce reflectors have been identified during
installation or one-time radio calibration. Their locations
$\{\vc q_s\}_{s=1}^{S}$ generate the RIS-domain matrix
\begin{equation}
\vc A_{\mathrm s}=
\begin{bmatrix}\vc a(\vc q_1)&\cdots&\vc a(\vc q_S)\end{bmatrix}.
\label{eq:scatter_atoms_paper}
\end{equation}
After division by the known pilot signal, the virtual RIS observation is
\begin{align}
\vc y_{\mathrm r}&=\vc h(\vc p)+\vc e_{\mathrm{mod}},\label{eq:yr_paper}\\
\vc h(\vc p)&=x_0\vc a(\vc p)+\vc A_{\mathrm s}\vc x_{\mathrm s}
=\vc D(\vc p)\vc x,\label{eq:h_sparse_paper}\\
\vc D(\vc p)&=\begin{bmatrix}\vc a(\vc p)&\vc A_{\mathrm s}\end{bmatrix},
\quad
\vc x=\begin{bmatrix}x_0&\vc x_{\mathrm s}^{\mathsf T}\end{bmatrix}^{\mathsf T}.
\label{eq:D_paper}
\end{align}
Here $\vc e_{\mathrm{mod}}$ collects unmodeled scattering and calibration
residuals. The nominal field has $S+1\ll N$ path components: $x_0$ is the
common LoS gain, and $x_s$ is the gain of reflected path $s$. The online
unknowns are the three coordinates of $\vc p$ and these $S+1$ complex gains,
not an independent parameter for every RIS element.

In particular, a fixed reflected-path atom does not imply a fixed reflected
signal. The vector $\vc a(\vc q_s)$ describes propagation from reflector $s$
to the RIS and is unchanged while that reflector remains stationary. Source
motion changes the distance and phase on the source-to-reflector segment; these
changes are absorbed into the snapshot-dependent coefficient $x_s$. Likewise,
$x_0$ absorbs the center-referenced LoS phase and attenuation, while
$\vc a(\vc p)$ captures their variation across elements. This separation is
what permits a calibrated reflected subspace to be reused without treating
instantaneous path gains as known.

To quantify departures from the nominal channel, the structural-mismatch experiment
instantiates $\vc e_{\mathrm{mod}}$ through
\begin{align}
\vc y_{\mathrm r}^{(\rho)}
&=\vc G(\rho)\big(\vc D(\vc p)\vc x
+\rho\vc A_{\mathrm u}\vc x_{\mathrm u}\big),
\label{eq:structural_mismatch_paper}\\
\vc G(\rho)
&=\diag\!\left((1+\rho\epsilon_n)e^{j\rho\psi_n}\right)_{n=1}^{N},
\quad 0\leq\rho\leq1,
\label{eq:element_response_paper}
\end{align}
where $\vc A_{\mathrm u}$ contains omitted weak paths and $\epsilon_n,\psi_n$
are repeatable element-response errors. OG-NF-SPR retains the nominal
$\vc D(\vc p)$ and $\vc G=\vc I$; learning can absorb these effects only when
the corresponding $\rho$ values are represented by labelled channels.
This construction separates two failures that otherwise look similar in an
NMSE curve. A metadata test perturbs only the reflector coordinates supplied
to the estimator while leaving the received samples unchanged; it measures the
cost of incorrect model side information. A structural test regenerates the
physical VOR through omitted paths and element-response errors; it measures
generalization to a channel family absent from the nominal model.

\subsection{Compressed BS Observation}

Let $\vc H\in\C^{M\times N}$ denote the RIS-to-BS channel and
$\vc\omega^{(\ell)}\in\C^N$ the $\ell$th predesigned RIS phase profile. The
source position and path gains are assumed constant while the $L$ profiles are
cycled within one snapshot. The $M$-antenna BS observation under profile
$\ell$ is
\begin{equation}
\vc y^{(\ell)}=\vc H\diag(\vc\omega^{(\ell)})\vc y_{\mathrm r}
+\vc n^{(\ell)},\quad \ell=1,\ldots,L.
\label{eq:profile_obs_paper}
\end{equation}
Stacking the $L$ profiles yields
\begin{align}
\vc y&=\vc\Phi\vc y_{\mathrm r}+\vc n
=\vc\Phi\vc D(\vc p)\vc x+\vc w,\label{eq:y_paper}\\
\vc\Phi&=\begin{bmatrix}
\vc H\diag(\vc\omega^{(1)})\\[-1mm]
\vdots\\[-1mm]
\vc H\diag(\vc\omega^{(L)})
\end{bmatrix}\in\C^{Q\times N},\quad Q=ML<N,
\label{eq:Phi_paper}\\
\vc w&=\vc\Phi\vc e_{\mathrm{mod}}+\vc n.
\label{eq:effective_noise_paper}
\end{align}
Here $|[\vc\omega^{(\ell)}]_n|=1$ and
$\vc n\sim\mathcal{CN}(\vc0,\vc R_n)$. When a covariance estimate is
available, the observation and sensing matrix are first left-whitened. We then
reuse $\vc y$ and $\vc\Phi$ for the whitened quantities and approximate the
effective disturbance in \eqref{eq:effective_noise_paper} as
$\vc w\sim\mathcal{CN}(\vc0,\sigma^2\vc I_Q)$. This approximation is exact for
white receiver noise and $\vc e_{\mathrm{mod}}=\vc0$; the mismatch experiment
later deliberately violates it. Direct inversion remains underdetermined
because $\vc\Phi$ has a nontrivial null space.

The information-theoretic boundary established for VOR in
\cite{WuJSTSP2026} remains essential here. Conditioned on the fixed deployment
mapping $g:\vc p\mapsto\vc y_{\mathrm r}$, the sensing chain is
$\vc p\to\vc y_{\mathrm r}\to\vc y$, and therefore
\begin{equation}
I(\vc p;\vc y_{\mathrm r}\mid g)
\geq I(\vc p;\vc y\mid g).
\label{eq:vor_information_paper}
\end{equation}
This inequality explains why the RIS-domain representation is desirable, but
not how an underdetermined inverse is selected. Once its side information is
fixed, either reconstructor is itself a post-processing of $\vc y$; hence,
\begin{align}
I(\vc p;\widehat{\vc y}_{\mathrm r}^{\mathrm L}\mid\mathcal D_{\mathrm{tr}})
&\leq I(\vc p;\vc y\mid\mathcal D_{\mathrm{tr}}),\nonumber\\
I(\vc p;\widehat{\vc y}_{\mathrm r}^{\mathrm M}\mid\mathcal I_{\mathrm M})
&\leq I(\vc p;\vc y\mid\mathcal I_{\mathrm M}).
\label{eq:reconstructor_information_paper}
\end{align}
Thus, neither branch creates online information. Learning uses paired
deployment samples as side information, whereas modeling uses calibrated
geometry and path statistics. If $g$ is injective and $\vc\Phi$ is injective
on the low-dimensional channel manifold, noiseless identification can still be
possible despite $Q<N$ \cite{WuJSTSP2026}; the present work studies which prior
approximates that inverse more reliably under noise, finite data, and mismatch.

The physical route thus replaces direct inversion by parameter estimation:
\begin{equation}
\underbrace{(\vc y,\vc\Phi)}_{\text{available at the BS}}
\longrightarrow (\widehat{\vc p},\widehat{\vc x})
\longrightarrow
\widehat{\vc y}_{\mathrm r}
=\vc D(\widehat{\vc p})\widehat{\vc x}.
\label{eq:inference_chain_paper}
\end{equation}
Here, geometry, path-power scales, noise variance, and the search region are
calibrated inputs; position and instantaneous gains remain unknown. Learning
bypasses these intermediate estimates and predicts the VOR from paired data.
The following two subsections specify the corresponding objectives.

\subsection{Modeling: Geometry-Constrained MAP Reconstruction}

Assume a zero-mean complex Gaussian path-gain prior
\begin{equation}
\vc x\sim\mathcal{CN}(\vc0,\vc R_x),\quad
\vc R_x=\diag(\nu_0,\nu_1,\ldots,\nu_S),
\label{eq:path_prior_paper}
\end{equation}
where $\nu_s$ specifies a path-power scale. With a uniform position prior over
$\mathcal P$, the joint MAP problem is
\begin{equation}
\min_{\vc p\in\mathcal P,\vc x}
\frac{1}{\sigma^2}\|\vc y-\vc\Phi\vc D(\vc p)\vc x\|_2^2
+\vc x^{\Hc}\vc R_x^{-1}\vc x.
\label{eq:map_problem_paper}
\end{equation}
The first term measures agreement with the received samples; the second
penalizes gains that are implausibly large relative to their calibrated
powers. The objective is quadratic in the $S+1$ complex gains and nonconvex
in only three real position variables. Its required side information is
\begin{equation}
\mathcal I_{\mathrm M}=\{\vc\Phi,\{\vc p_n\},\vc p_{\mathrm R},
\{\vc q_s\},\vc R_x,\sigma^2,\mathcal P\},
\label{eq:model_information_paper}
\end{equation}
comprising the sensing operator, array and reflector geometry, gain-power
scales, noise level, and search region. No instantaneous gain is included.

The Gaussian prior supplies regularization, rather than asserting that
physical gains cannot vary with source position. For every candidate
position, the gains are fitted anew to the observation. Consequently, the
high-dimensional output is reconstructed through a small set of physical
parameters. This reduction is useful only to the extent that the assumed
channel family describes the deployment.

\subsection{Learning: Supervised VOR Reconstruction}

The learned route estimates the aperture signal without fitting an explicit
source location. Let
$\vc z=[\RePart\{\vc y\}^{\mathsf T},\Im\{\vc y\}^{\mathsf T}]^{\mathsf T}$
and define $\vc t$ analogously from $\vc y_{\mathrm r}$. A training pair
$(\vc z,\vc t)$ therefore contains a BS input and its clean RIS target,
not a position label. Sample splitting and noise augmentation are specified
in Section~\ref{sec:results}.

Let $\overline{\vc z}=\vc S_z(\vc z-\vc\mu_z)$ and
$\overline{\vc t}=\vc S_t(\vc t-\vc\mu_t)$, where the means and diagonal
inverse-standard-deviation matrices are fitted only on training pairs. Both
networks minimize
\begin{equation}
\widehat{\vc\theta}_m=\arg\min_{\vc\theta_m}
\frac{1}{N_{\mathrm{tr}}}\sum_{d=1}^{N_{\mathrm{tr}}}
\|\overline{\vc t}_d-f_m(\overline{\vc z}_d;\vc\theta_m)\|_2^2,
\label{eq:learning_objective_paper}
\end{equation}
where $m\in\{\mathrm{DNN},\mathrm{Dense}\}$ and barred variables are
standardized real representations. Compact VOR-DNN is an ordering-agnostic
control with two width-512 hidden layers:
\begin{equation}
f_{\mathrm{DNN}}(\overline{\vc z})
=\vc W_3\rho\big(\vc W_2\rho(\vc W_1\overline{\vc z}+\vc b_1)
+\vc b_2\big)+\vc b_3.
\label{eq:compact_dnn_paper}
\end{equation}
The layers use normalization, GELU activation $\rho(\cdot)$, and dropout.
The DenseNet-based reconstructor, termed DenseNet-VOR, reshapes the input to
the fixed $2\times4\times4$ measurement layout
and applies DenseNet connectivity
\begin{equation}
\vc F_{b,\ell}=\mathcal H_{b,\ell}
\big([\vc F_{b,0},\ldots,\vc F_{b,\ell-1}]\big),
\label{eq:dense_connectivity_paper}
\end{equation}
where $[\cdot]$ concatenates channels and $\mathcal H_{b,\ell}$ is a
BN--ReLU--$3\times3$ convolution. Three four-layer blocks use growth rates 16,
20, and 24, with $1\times1$ transitions, average pooling, global pooling, and
a width-512 output head. It preserves the feature reuse and BS-to-VOR objective of
\cite{WuJSTSP2026} under the present $Q=16$ layout and label budget; Compact
VOR-DNN controls for this local convolutional bias.

At inference, the stored preprocessing is reused and the network output is
converted back through
$\widehat{\vc t}=\vc\mu_t+\vc S_t^{-1}
f_m(\overline{\vc z};\widehat{\vc\theta}_m)$ before assembling the complex
VOR. A fixed measurement ordering is maintained throughout training and testing.
The $4\times4$ input arrangement is a computational layout of compressed
measurements, not a directly sampled patch of the RIS. Consequently, the
DenseNet architecture is evaluated as a feature-reuse prior for this layout;
its convolutional structure alone does not establish generalization to another
sensing operator. Compact VOR-DNN does not impose the same local adjacency,
although it also requires a consistent input ordering.

We use two network architectures to distinguish learned reconstruction from
the effect of a particular feature extractor. A third estimator,
DR-VOR-DNN, keeps the Compact architecture and parameter count but changes
the training channels to include the mismatch in
\eqref{eq:structural_mismatch_paper}. It tests training coverage rather than
additional network capacity. All networks retain fixed preprocessing and
weights during testing; they do not adapt to individual test samples.

\subsection{Common Target, Different Prior Information}

Both routes return the same type of estimate:
\begin{align}
\widehat{\vc y}_{\mathrm r}^{\mathrm M}
&=f_{\mathrm M}(\vc y;\mathcal I_{\mathrm M}),\nonumber\\
\widehat{\vc y}_{\mathrm r}^{\mathrm L}
&=f_{\mathrm L}(\vc y;\widehat{\vc\theta},
\mathcal D_{\mathrm{tr}}).
\label{eq:two_priors_paper}
\end{align}
Here $\mathcal D_{\mathrm{tr}}$ specifies the paired training data, including
the preprocessing learned from it; these data need not be queried during
network inference. Modeling instead uses $\mathcal I_{\mathrm M}$ at each
snapshot. Neither route is prior-free: one requires calibration, and the
other requires representative labels.

\begin{figure*}[t]
\centering
\begin{tikzpicture}[
font=\rmfamily\footnotesize,
flow/.style={-{Latex[length=1.5mm,width=1mm]},line width=0.55pt},
sideinfo/.style={flow,dash pattern=on 2.5pt off 2pt},
block/.style={draw=black,line width=0.55pt,align=center,
  minimum height=9mm,inner sep=3pt},
signal/.style={inner sep=1.5pt},
labeltext/.style={fill=white,inner sep=1.5pt}
]
\node[signal] (yl) at (0,0) {$\vc y$};
\node[block,text width=25mm] (scale) at (2.5,0)
{Input\\standardization};
\node[block,text width=43mm] (network) at (7.5,0)
{Frozen VOR network\\$f_m(\overline{\vc z};\widehat{\vc\theta}_m)$};
\node[block,text width=25mm] (assemble) at (12.5,0)
{Rescaling and\\complex assembly};
\node[signal] (rl) at (15.1,0)
{$\widehat{\vc y}_{\mathrm r}^{\mathrm L}$};

\draw[flow] (yl) -- (scale);
\draw[flow] (scale) -- (network)
node[midway,above] {$\overline{\vc z}$};
\draw[flow] (network) -- (assemble)
node[midway,above] {$\widehat{\overline{\vc t}}$};
\draw[flow] (assemble) -- (rl);

\node[block,minimum height=7mm,text width=60mm] (training) at (7.5,1.5)
{Paired BS/RIS data: offline training};
\draw[sideinfo] (training) -- (network)
node[midway,right] {$\widehat{\vc\theta}_m$};
\node at (7.5,-0.9) {(a) Learning-based reconstruction};

\node at (7.5,-1.75)
{Given $\mathcal I_{\mathrm M}$: sensing operator, geometry, and path/noise statistics};
\node[signal] (ym) at (0,-3.2) {$\vc y$};
\node[block,text width=25mm] (coarse) at (2.5,-3.2)
{Projection and\\coarse search};
\node[block,text width=43mm] (refine) at (7.5,-3.2)
{MAP gain estimation\\and off-grid refinement};
\node[block,text width=25mm] (synthesis) at (12.5,-3.2)
{RIS synthesis\\$\vc D(\widehat{\vc p})\widehat{\vc x}$};
\node[signal] (rm) at (15.1,-3.2)
{$\widehat{\vc y}_{\mathrm r}^{\mathrm M}$};

\draw[flow] (ym) -- (coarse);
\draw[flow] (coarse) -- (refine)
node[midway,above] {seeds};
\draw[flow] (refine) -- (synthesis)
node[midway,above] {$\widehat{\vc p},\widehat{\vc x}$};
\draw[flow] (synthesis) -- (rm);
\fill (0.65,-3.2) circle (1pt);
\draw[flow] (0.65,-3.2) -- (0.65,-2.35) -| (refine.north)
node[pos=0.45,above] {$\vc y$};
\draw[flow] ([xshift=11mm]refine.south) -- (8.6,-4.15)
-- (6.4,-4.15) -- ([xshift=-11mm]refine.south);
\node[labeltext] at (7.5,-4.15) {iterate};
\node at (7.5,-4.75) {(b) Model-based reconstruction (OG-NF-SPR)};
\end{tikzpicture}
\caption{Two reconstruction routes for the same compressed observation
$\vc y=\vc\Phi\vc y_{\mathrm r}+\vc n$, $Q<N$.
(a) Offline-trained VOR inference with fixed preprocessing.
(b) Geometry-constrained estimation followed by RIS synthesis; the full
observation bypasses the initialization stage. The dashed arrow supplies
trained weights; solid arrows indicate online data flow. Either output may
feed the optional SSFE--BiTP tracking chain.}
\label{fig:learning_modeling_framework}
\end{figure*}
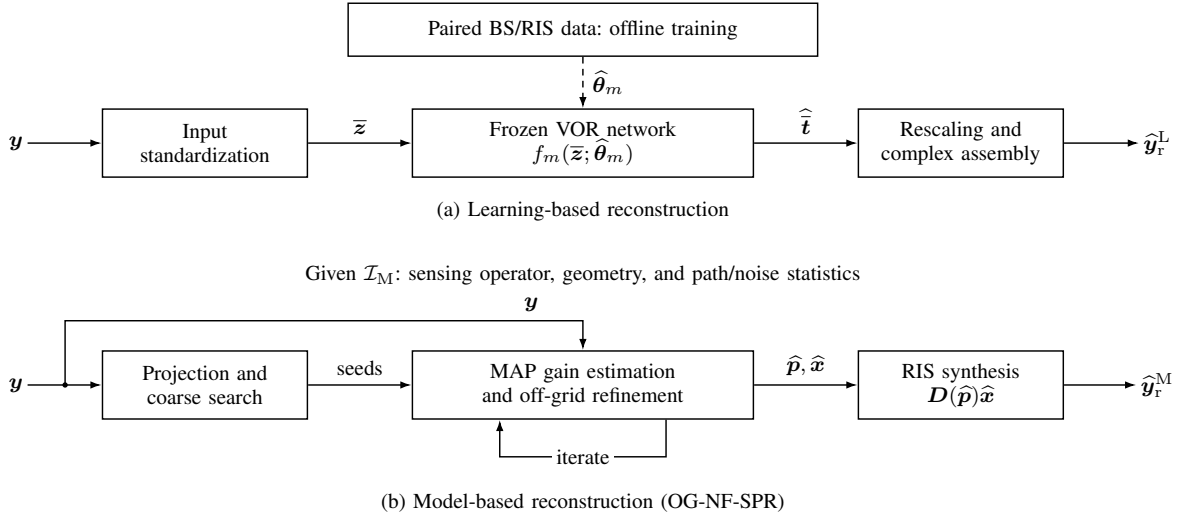

Fig.~\ref{fig:learning_modeling_framework} shows the resulting computations.
In the upper route, fixed preprocessing and a frozen network produce the
VOR in one forward pass. In the lower route, projection initializes position
seeds, but the full observation is used to fit all gains and refine position
before RIS synthesis. The feedback loop represents per-snapshot fitting, not
offline training. Whether the computational and data requirements are
worthwhile depends on the reliability of the available prior, which the
experiments examine separately.

\subsection{Deep VOR-Assisted Tracking Formulation}

For tracking, either reconstructed sequence can feed the SSFE--BiTP chain
\begin{equation}
\widehat{\vc p}_t=\mathcal G_{\mathrm{BiTP}}\!
\left(\{\mathcal G_{\mathrm{SSFE}}(\widehat{\vc y}_{\mathrm r,\tau})
\}_{\tau=t-T+1}^{t}\right).
\label{eq:deep_tracking_paper}
\end{equation}
where the spatial-spectral feature extractor (SSFE) extracts spatial/angular
features and the bidirectional LSTM predictor (BiTP) maps the available
length-$T$ window to the current 3D position \cite{WuJSTSP2026}. The shared-trajectory reconstruction
experiment uses causal T-OG-NF-SPR to exploit a motion prior without supervised
trajectory labels.

\section{Modeling Algorithm: OG-NF-SPR}\label{sec:algorithm}

The main difficulty in \eqref{eq:map_problem_paper} is the coupling between
source position and unknown path gains. For a fixed position, however, the
gain-estimation subproblem is quadratic and has a closed-form solution.
OG-NF-SPR exploits this structure to replace joint nonlinear fitting by a
three-dimensional position search with gain re-estimation at each trial.

The remaining challenges are initialization and finite-grid resolution.
Reflected-path projection makes the initial search less dependent on
instantaneous multipath gains, while multiple seeds reduce sensitivity to an
incorrect local maximum. Continuous refinement then removes the restriction
to grid locations. Crucially, projection is used only for initialization:
the full observation and all calibrated paths enter the subsequent fit.

\subsection{Initialization by Multipath Projection}

Directly correlating $\vc y$ with a candidate LoS atom can favor a reflected
component rather than the source. We first remove the subspace generated by
the calibrated reflectors. Define the compressed reflected-path matrix and
its orthogonal-complement projector as
\begin{equation}
\vc B_{\mathrm s}=\vc\Phi\vc A_{\mathrm s},\quad
\vc P_{\mathrm s}^{\perp}=\vc I_Q-\vc B_{\mathrm s}\vc B_{\mathrm s}^{\dagger}.
\label{eq:scatter_projector_paper}
\end{equation}
The projector is formed by a rank-revealing QR or SVD in implementation, which
avoids explicitly evaluating the pseudoinverse when $\vc B_{\mathrm s}$ is
ill-conditioned. If no calibrated reflector is retained, then
$\vc P_{\mathrm s}^{\perp}=\vc I_Q$.
Since $\vc P_{\mathrm s}^{\perp}\vc B_{\mathrm s}=\vc0$, multiplication by
$\vc P_{\mathrm s}^{\perp}$ suppresses components that can be explained by
the fixed reflector atoms without requiring their unknown instantaneous
gains. Hence, the projected observation satisfies
\begin{equation}
\widetilde{\vc y}=\vc P_{\mathrm s}^{\perp}\vc y
\approx x_0\vc P_{\mathrm s}^{\perp}\vc\Phi\vc a(\vc p)
+\widetilde{\vc n}.
\label{eq:projected_y_paper}
\end{equation}
For a three-dimensional grid $\mathcal G=\{\vc p_g\}_{g=1}^{G}$, precompute
\begin{equation}
\widetilde{\vc\psi}_g=
\frac{\vc P_{\mathrm s}^{\perp}\vc\Phi\vc a(\vc p_g)}
{\|\vc P_{\mathrm s}^{\perp}\vc\Phi\vc a(\vc p_g)\|_2}.
\label{eq:normalized_atom_paper}
\end{equation}
The coarse seed is selected using
\begin{equation}
g^{(0)}=\arg\max_g
|\widetilde{\vc\psi}_g^{\Hc}\widetilde{\vc y}|,\quad
\vc p^{(0)}=\vc p_{g^{(0)}}.
\label{eq:coarse_paper}
\end{equation}
Equation~\eqref{eq:coarse_paper} is a matched-filter search over candidate source
positions after reflected-subspace cancellation. The implementation retains
$C$ spatially separated high-score seeds to reduce local-minimum failures.
Normalizing each atom makes the score measure directional agreement rather
than received energy, thereby avoiding preference for a grid point merely
because its compressed steering vector has larger norm.

\subsection{Closed-Form Complex Path-Gain Estimation}

For a fixed $\vc p$, let
\begin{equation}
\vc B(\vc p)=\vc\Phi\vc D(\vc p)
=\begin{bmatrix}\vc\Phi\vc a(\vc p)&\vc B_{\mathrm s}\end{bmatrix}.
\label{eq:B_paper}
\end{equation}
Once $\vc p$ is fixed, the nonlinear problem becomes a regularized linear
regression in the $S+1$ complex gains. Setting the derivative of
\eqref{eq:map_problem_paper} with respect to $\vc x^*$ to zero gives the normal
equation
\begin{equation}
\big(\vc B^{\Hc}\vc B+\sigma^2\vc R_x^{-1}\big)\widehat{\vc x}
=\vc B^{\Hc}\vc y,
\label{eq:normal_gain_paper}
\end{equation}
and therefore the conditional MAP estimate
\begin{equation}
\widehat{\vc x}(\vc p)=
\big(\vc B^{\Hc}\vc B+\sigma^2\vc R_x^{-1}\big)^{-1}
\vc B^{\Hc}\vc y,
\label{eq:xmap_paper}
\end{equation}
where the argument $\vc p$ on $\vc B$ is omitted on the right for compactness.
Substitution into \eqref{eq:map_problem_paper} gives the profile objective
\begin{equation}
F(\vc p)=\|\vc y-\vc B(\vc p)\widehat{\vc x}(\vc p)\|_2^2
+\sigma^2\widehat{\vc x}^{\Hc}(\vc p)
\vc R_x^{-1}\widehat{\vc x}(\vc p).
\label{eq:profile_paper}
\end{equation}
The function $F(\vc p)$ scores a candidate position only after fitting all
gains at that position. Thus, gain differences do not have to be explained
by moving the source. The regularizer limits unstable gains at low SNR.
Importantly, $\vc B(\vc p)$ contains both LoS and reflected atoms: the
reflection components removed for initialization now contribute to the fit
and the final reconstruction.

\subsection{Continuous Off-Grid Refinement}

A grid point is an initial guess, not a restriction on the final position.
We refine it by asking how a small displacement changes the residual after
the path gains have been fitted. Let $k=2\pi/\lambda$. Differentiating \eqref{eq:atom_paper} gives
\begin{align}
\frac{\partial[\vc a(\vc u)]_n}{\partial u_i}
&=[\vc a(\vc u)]_n\xi_{n,i}(\vc u),\label{eq:atom_derivative_paper}\\
\xi_{n,i}(\vc u)
&=\frac{u_i-p_{\mathrm R,i}}{d_c^2(\vc u)}
-\frac{u_i-p_{n,i}}{d_n^2(\vc u)}\nonumber\\
&\quad-jk\left(\frac{u_i-p_{n,i}}{d_n(\vc u)}
-\frac{u_i-p_{\mathrm R,i}}{d_c(\vc u)}\right).
\label{eq:log_derivative_paper}
\end{align}
Collect the three derivatives in $\vc D_a(\vc p)\in\C^{N\times3}$. At
iteration $t$, only the source-dependent LoS atom is differentiated because
$\vc A_{\mathrm s}$ is fixed. Define
\begin{align}
\vc r^{(t)}&=\vc y-\vc B(\vc p^{(t)})
\widehat{\vc x}(\vc p^{(t)}),\label{eq:residual_paper}\\
\vc P_B^{\perp}&=\vc I_Q-\vc B(\vc p^{(t)})
\vc B^{\dagger}(\vc p^{(t)}),\label{eq:PB_paper}\\
\vc J^{(t)}&=\vc P_B^{\perp}\vc\Phi\vc D_a(\vc p^{(t)})
\widehat{x}_0(\vc p^{(t)}).
\label{eq:J_paper}
\end{align}
Equation \eqref{eq:J_paper} is the projected-residual Jacobian: it removes
directions that can be explained by re-estimating the complex gains. Because
$\widehat{\vc x}$ is ridge regularized, this is a stable variable-projection
approximation rather than the exact Hessian of \eqref{eq:profile_paper}; every
trial is therefore checked against the exact profiled MAP objective. Locally,
the residual is approximated by
\begin{equation}
\vc r(\vc p^{(t)}+\Delta\vc p)
\simeq \vc r^{(t)}-\vc J^{(t)}\Delta\vc p,
\label{eq:residual_linearization_paper}
\end{equation}
where $\Delta\vc p\in\R^3$. Minimizing this real-parameter least-squares model
with damping $\mu_t$ gives
\begin{align}
\Delta\vc p^{(t)}
&=\left(\RePart\{(\vc J^{(t)})^{\Hc}\vc J^{(t)}\}
+\mu_t\vc I_3\right)^{-1}\nonumber\\
&\quad\times\RePart\{(\vc J^{(t)})^{\Hc}\vc r^{(t)}\}.
\label{eq:GN_paper}
\end{align}
The step is clipped to a trust radius, projected onto $\mathcal P$, and
accepted only when the gains are recomputed and
$F(\vc p^{(t+1)})<F(\vc p^{(t)})$. Otherwise, the current point is retained and
the trust radius is halved. This safeguard separates guaranteed objective
descent from the local quadratic approximation. After $T_{\mathrm{GN}}$
iterations,
\begin{equation}
\widehat{\vc y}_{\mathrm r}
=\vc D(\widehat{\vc p})\widehat{\vc x}(\widehat{\vc p}).
\label{eq:recon_paper}
\end{equation}
The output of \eqref{eq:recon_paper} is the $N$-element aperture signal, not
just the estimated LoS path. Position is an intermediate variable that also
provides a localization output; reconstruction and position errors will
therefore be evaluated separately.

\subsection{Trajectory-Aware Extension}

Independent snapshots can select different candidate locations when several
fit the data nearly equally well. A previous estimate supplies additional
information for resolving this ambiguity. Consider the random-walk prior
\begin{equation}
\vc p_t=\vc p_{t-1}+\vc v_t,\quad
\vc v_t\sim\mathcal N(\vc0,\vc Q_p),\quad
\vc p_t^-=\widehat{\vc p}_{t-1}.
\label{eq:motion_paper}
\end{equation}
Here $\vc p_t^-$ is the one-step position prediction and $\vc Q_p$ controls
how strongly the current estimate is tied to it: a small $\vc Q_p$ favors a
smooth trajectory, whereas a large $\vc Q_p$ approaches independent-snapshot
reconstruction and permits rapid maneuvers.
The causal temporal MAP profile objective is
\begin{equation}
F_t^{\mathrm T}(\vc p)=F_t(\vc p)
+\sigma^2\|\vc p-\vc p_t^-\|_{\vc Q_p^{-1}}^2.
\label{eq:temporal_profile_paper}
\end{equation}
For $\vc Q_p=\sigma_p^2\vc I_3$, the coarse score is heuristically biased
toward the predicted basin using
\begin{equation}
s_{t,g}^{\mathrm T}
=\log\left(\frac{|\widetilde{\vc\psi}_g^{\Hc}\widetilde{\vc y}_t|}
{\max_i|\widetilde{\vc\psi}_i^{\Hc}\widetilde{\vc y}_t|}+\epsilon\right)
-\frac{\|\vc p_g-\vc p_t^-\|_2^2}{2\sigma_p^2}.
\label{eq:temporal_score_paper}
\end{equation}
The logarithm makes the correlation term scale invariant; final seed selection
still uses the exact objective in \eqref{eq:temporal_profile_paper}. With
$\vc\Lambda_p=\sigma^2\vc Q_p^{-1}$, the normal matrix in
\eqref{eq:GN_paper} gains $\vc\Lambda_p$, while its right-hand side gains
$-\vc\Lambda_p(\vc p_t^{(i)}-\vc p_t^-)$. This implementation is denoted by
T-OG-NF-SPR. Algorithm~\ref{alg:ognfspr} presents the complete procedures for
both OG-NF-SPR and T-OG-NF-SPR.

\begin{algorithm}[H]
\caption{OG-NF-SPR/T-OG-NF-SPR Reconstruction}
\label{alg:ognfspr}
\begin{algorithmic}[1]
\REQUIRE $\vc y$, $\vc\Phi$, $\vc A_{\mathrm s}$, $\mathcal G$,
$\vc R_x$, $\sigma^2$, and optional $(\vc p_t^-,\vc Q_p)$.
\ENSURE Reconstructed $\widehat{\vc y}_{\mathrm r}$ and position
$\widehat{\vc p}$.
\STATE Form $\vc B_{\mathrm s}$ and $\vc P_{\mathrm s}^{\perp}$; compute
$\widetilde{\vc y}=\vc P_{\mathrm s}^{\perp}\vc y$.
\STATE Score every $\vc p_g\in\mathcal G$ by
\eqref{eq:coarse_paper}, or by \eqref{eq:temporal_score_paper} when a motion
prior is available.
\STATE Retain $C$ spatially separated coarse seeds.
\FOR{each retained seed}
\STATE Initialize $\vc p^{(0)}$ at that grid point and its trust radius.
\FOR{$i=0,\ldots,T_{\mathrm{GN}}-1$}
\STATE Estimate $\widehat{\vc x}(\vc p^{(i)})$ by
\eqref{eq:xmap_paper}.
\STATE Compute $\vc D_a$, $\vc r^{(i)}$, and $\vc J^{(i)}$.
\STATE Form, clip, and region-project the independent or temporal
Gauss--Newton trial point.
\STATE Re-estimate its gains; accept it only if the exact profile objective
decreases, and otherwise shrink the trust radius.
\ENDFOR
\ENDFOR
\STATE Re-estimate all path gains and select the seed with the smallest exact
profile objective.
\STATE Return $\widehat{\vc y}_{\mathrm r}$ from
\eqref{eq:recon_paper} and $\widehat{\vc p}$.
\end{algorithmic}
\end{algorithm}

\section{Analysis and Interpretation}\label{sec:analysis}

The analysis addresses two different questions: whether refinement decreases
its objective, and whether nearby source positions can be distinguished.
Objective descent alone does not imply accurate localization.

\subsection{Descent and Local Identifiability}

\begin{proposition}[Monotonic accepted descent]
For each retained seed, the sequence of objective values accepted by
OG-NF-SPR is nonincreasing and convergent. The same statement holds for
T-OG-NF-SPR with $F_t^{\mathrm T}$.
\end{proposition}
\begin{proof}
Accepted trials strictly decrease the nonnegative profile objective; rejected
trials retain its previous value. The sequence therefore converges, although
nonconvexity precludes a global-optimality guarantee.
\end{proof}

\begin{proposition}[Local geometric identifiability]
After nuisance-subspace removal, the position is locally identifiable at
$\vc p$ if
\begin{equation}
\|\vc P_{\mathrm s}^{\perp}\vc\Phi\vc a(\vc p)\|_2>0,
\quad
\operatorname{rank}(\RePart\{\vc J^{\Hc}\vc J\})=3.
\label{eq:identifiability_paper}
\end{equation}
\end{proposition}
\begin{proof}
The first condition preserves the LoS atom after nuisance projection, and the
rank condition makes the first-order map from three position perturbations to
the residual injective. Local uniqueness follows from the inverse-function
argument.
\end{proof}

The two conditions have separate meanings: projection must retain a nonzero
LoS component, and displacements along all three coordinates must produce
distinguishable residual changes. Near-field curvature supplies radial
sensitivity, but compression can weaken it. Under the matched white-noise
model, a local Fisher-information approximation and its covariance benchmark
for locally unbiased estimation are
\begin{equation}
\vc J_{\vc p}\approx\frac{2}{\sigma^2}\RePart\{\vc J^{\Hc}\vc J\},
\quad
\operatorname{Cov}(\widehat{\vc p})\succeq\vc J_{\vc p}^{-1}.
\label{eq:local_crb_paper}
\end{equation}
This benchmark describes small errors around the true position. It is not a
bound asserted for the biased regularized estimator under mismatch, nor does
it account for selecting an incorrect coarse location at low SNR.

\subsection{Why Learning and Modeling Cross Over}

For squared-error training and unlimited representative data, the population
minimizer is the conditional mean
\begin{equation}
f_{\mathrm L}^{\star}(\vc z)
=\mathbb E_{\mathrm{tr}}[\vc t\mid\vc z].
\label{eq:conditional_mean_paper}
\end{equation}
The conditional mean and OG-NF-SPR make different decisions under uncertainty.
If several positions remain plausible, the former averages their aperture
signals, whereas the latter selects a position and fitted gains. Averaging
can reduce squared error even when the average is not the signal of any
single source location. This explains why reconstruction accuracy and
localization accuracy need not improve together.

The averaging advantage is not exclusive to deep learning: LMMSE also
uses deployment statistics, and a model-based posterior-mean estimator would
differ from the MAP procedure studied here. Finite networks only approximate
\eqref{eq:conditional_mean_paper}, with accuracy tied to the training
distribution. The low-SNR comparison therefore concerns the evaluated
estimators, not an inherent advantage of learning over all Bayesian methods.

With correct calibration and informative observations, OG-NF-SPR can
distinguish nearby positions and reduce grid error through refinement.
Its model remains applicable to positions absent from the training set as
long as they lie in $\mathcal P$. Incorrect geometry instead changes the
admissible signal family; more accurate optimization cannot, by itself,
correct that family. Mismatch-aware training offers another way to describe
a repeatable discrepancy, provided that it is represented in the labels.
These mechanisms motivate the separate noise, coverage, and mismatch tests
in Section~\ref{sec:results}.

\subsection{Computational Complexity}

Let $K=S+1$. With precomputed projected grid atoms, the online complexity per
snapshot is
\begin{equation}
\mathcal O(QG)+\mathcal O\left(CT_{\mathrm{GN}}
(QN+QK^2+K^3)\right).
\label{eq:complexity_paper}
\end{equation}
The parallelizable $QG$ search dominates for a large grid; T-OG-NF-SPR adds
only an $\mathcal O(G)$ distance penalty, and storage is $\mathcal O(QG)$.

For Compact VOR-DNN with input dimension $2Q$, output dimension $2N$, and
hidden width $H$, inference requires
\begin{equation}
\mathcal O(2QH+H^2+2NH)
\label{eq:dnn_complexity_paper}
\end{equation}
multiply--accumulates; DenseNet-VOR uses dense-block convolutions instead.
The networks contain 0.384M/0.415M parameters and measured 0.73/4.92 $\mu$s
CUDA batch latency. OG-NF-SPR requires 0.214--1.419 ms in MATLAB on CPU as
refinement increases from zero to eight iterations. Hardware and batching
differ, so these timings do not establish a platform-independent speedup.
Neural inference has a fixed forward cost plus offline storage and training;
physical fitting repeats search and refinement per snapshot. DR-VOR-DNN
has the Compact network's size.

\section{Simulation Results: Learning or Modeling?}\label{sec:results}

We illustrate the comparison using near-field UAV tracking under a fixed
measurement operator. The experiments address three issues: accuracy when
the nominal model is valid, sensitivity when either calibration or
propagation changes, and the benefit of temporal information. Within each
reconstruction test, all methods receive the same signal and noise
realizations. The final tracking ablation uses the separate protocol of
\cite{WuJSTSP2026} to explain the role of the reconstructed representation.

\begin{figure*}[t]
\centering
\includegraphics[width=0.98\textwidth]{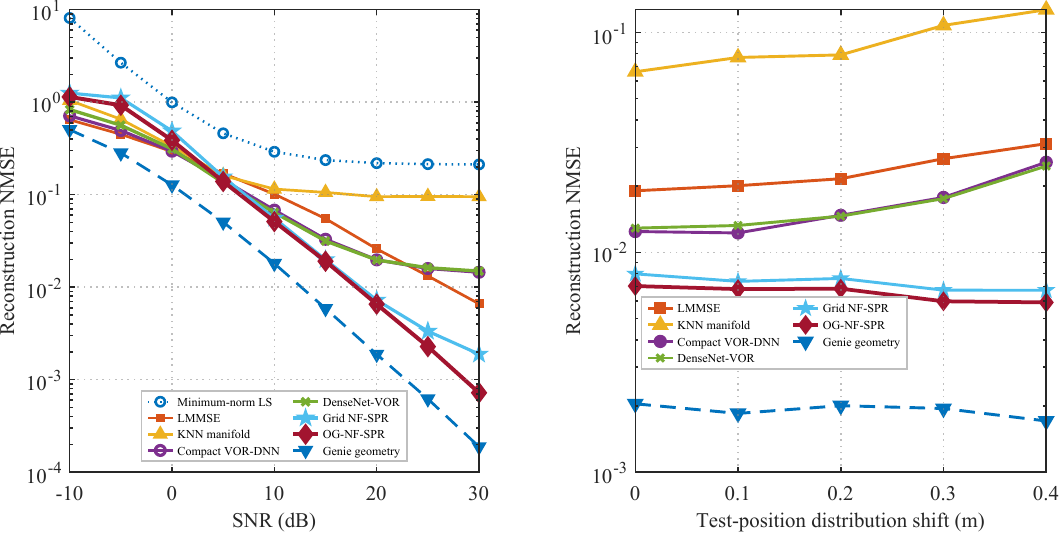}
\caption{VOR reconstruction performance. Left: NMSE versus SNR under the
matched position distribution. Right: NMSE at 20 dB under a range-direction
test-position shift while both deep models remain fixed.}
\label{fig:snr_paper}
\end{figure*}

\subsection{Simulation Setup}

We consider a $10\times10$ half-wavelength UPA operating at 3.5 GHz, with
$N=100$ RIS elements and $Q=16$ compressed measurements. The RIS lies in the
$y$--$z$ plane with its center at the origin. Nominal UAV positions follow a
Gaussian distribution centered at $[2,2,1.2]^{\mathsf T}$ m with 0.18 m
coordinate-wise standard deviation. Two dominant reflectors are located at
$[1.05,-1.10,1.75]^{\mathsf T}$ and $[1.40,3.35,0.55]^{\mathsf T}$ m, with
relative path magnitudes 0.34 and 0.22. Continuous test positions avoid
restricting the evaluation to the estimator's grid. Unless specified otherwise,
the results use 400 held-out snapshots.

The effective BS operator is a row-orthonormal basis constructed from the
nominal channel's three position derivatives and additional phase-code
columns. This idealized construction preserves local geometric sensitivity,
and all reconstructors use the same fixed $\vc\Phi$. It is not a demonstration
that an arbitrary such operator is realizable by a given $\vc H$ and passive
phase codebook in \eqref{eq:Phi_paper}. The comparison concerns inference under
calibrated compression; physical probing design is not optimized here.

OG-NF-SPR searches a $17^3$ grid over a 1.3 m cube, retains three seeds, and
performs six refinements. Data-dependent baselines share 5000 labelled
channels generated offline from the simulated deployment. The cost of
acquiring real RIS labels is not evaluated. For deep training, these channels are split into 4500/500
training/validation channels before adding noise. Each training channel
provides one clean BS input and two independently corrupted inputs with SNRs
uniformly drawn from $[-10,30]$ dB; each validation channel provides one clean
and one noisy input. All copies share their channel's clean RIS target,
yielding 13500/1000 pairs with no underlying target shared between splits.

Training uses AdamW, initial learning rate $10^{-3}$, batch size 256, and at
most 35 epochs with validation early stopping. The lowest-validation-loss
checkpoint is fixed during testing. DR-VOR-DNN uses the same 5000 source
positions, train/validation split, and noise augmentation, but generates each
channel with an independently drawn $\rho\sim\mathcal U[0,1]$ in
\eqref{eq:structural_mismatch_paper}. Thus, it changes channel coverage without
increasing the label count or network size. Table~\ref{tab:parameters_paper}
summarizes the parameters.

\begin{table}[t]
\centering
\caption{Main Simulation Parameters}
\label{tab:parameters_paper}
\begin{tabular}{l|c}
\toprule
Parameter & Value \\
\midrule
Carrier frequency & 3.5 GHz \\
RIS size / compressed dimension & $N=100$ / $Q=16$ \\
Coarse grid / dominant reflectors & $17^3$ / $S=2$ \\
Retained seeds / refinements & $C=3$ / $T_{\mathrm{GN}}=6$ \\
Test snapshots & 400 \\
Labelled channels & 5000 (4500/500 split) \\
Augmented train/validation pairs & 13500 / 1000 \\
Deep-model parameter counts & 0.384M / 0.415M \\
\bottomrule
\end{tabular}
\end{table}

The benchmarks answer complementary questions. Minimum-norm LS tests
inversion without a channel prior; LMMSE uses the offline mean and covariance;
and KNN retrieves the RIS target paired with the nearest stored BS sample.
Compact VOR-DNN and DenseNet-VOR test nonlinear regression with different
architectures. Grid NF-SPR uses the same physical model without continuous
refinement, isolating grid error. The known-position genie fits path gains
at the true UAV position, isolating the cost of estimating position. It is
an informed reference, not a universal error bound. All methods in each
reconstruction comparison use identical test channels and noise realizations.

For estimator $a$, reconstruction accuracy is measured by
\begin{equation}
\operatorname{NMSE}_a=
\frac{\sum_{i=1}^{N_{\mathrm{te}}}
\|\widehat{\vc y}_{\mathrm r,i}^{(a)}-\vc y_{\mathrm r,i}\|_2^2}
{\sum_{i=1}^{N_{\mathrm{te}}}\|\vc y_{\mathrm r,i}\|_2^2},
\label{eq:nmse_metric_paper}
\end{equation}
whereas position accuracy is reported separately as 3D RMSE.
SNR is the average clean BS sample power divided by the per-entry noise
variance, with the latter set for each test condition. The position-shift
sweep uses 200 test locations per shift, with 0.12 m coordinate-wise standard
deviation, means displaced along $x$, and coordinates clipped to the fixed
search box. Its zero-shift point thus need not equal the corresponding
400-snapshot SNR result. The downstream tracking ablation uses a separate
sequence protocol, specified with its results below.

\begin{figure*}[t]
\centering
\includegraphics[width=0.94\textwidth]{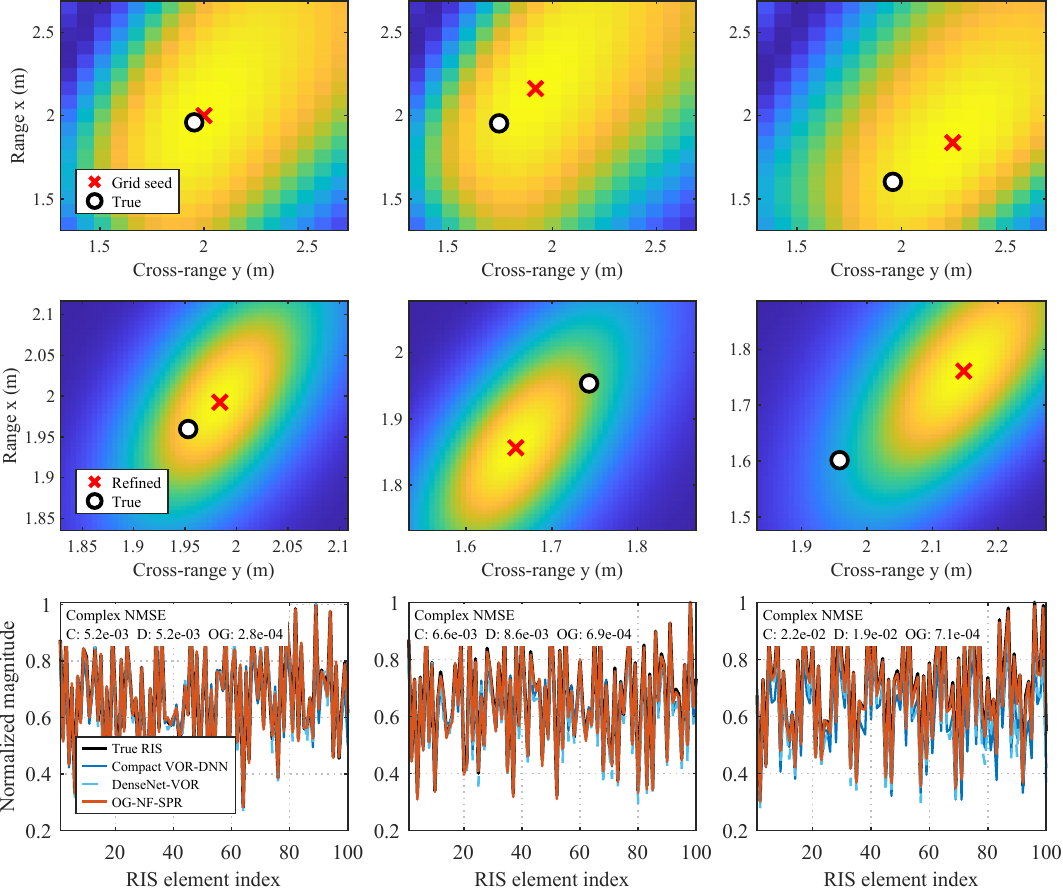}
\caption{Shared-snapshot estimates at 30 dB. Top/middle: projected coarse score
and refined MAP profile (red cross: estimate; black-outlined white circle: truth). Bottom:
per-element VOR magnitude for the truth and three reconstructors; in-panel
values are complex NMSEs (C: Compact, D: DenseNet-VOR, OG: OG-NF-SPR).}
\label{fig:surfaces_paper}
\end{figure*}

\subsection{Effect of SNR and Position-Distribution Shift}

Fig.~\ref{fig:snr_paper} compares noise sensitivity and position coverage.
In the left panel, both the test channel and the physical estimator use the
same spherical-wave family. This isolates estimation and approximation
errors before introducing model mismatch.

At $-5$ dB, Compact VOR-DNN attains 0.494 NMSE compared with 0.926 for
OG-NF-SPR, while LMMSE performs still better at 0.450. When several geometric
supports explain a noisy observation, averaging over the deployment
distribution can be more effective than selecting one support. This benefit
therefore does not require a deep network. As the observations become more informative, DenseNet-VOR and OG-NF-SPR
are comparable at 5 dB (0.135 and 0.139), and OG-NF-SPR is better by 10 dB.
At 20 dB, its NMSE is $6.553\times10^{-3}$, approximately 67\% below the
$1.97\times10^{-2}$ achieved by either deep network.

The baselines identify different sources of residual error. Minimum-norm LS
cannot recover components removed by compression, even when noise is small.
KNN is limited by single-sample retrieval, whereas the deep networks learn
a regression between samples. Their similar high-SNR plateaus reflect the
tested architectures and training budget rather than an intrinsic limit of
deep learning; LMMSE also falls below both networks at 30 dB.

For the physical estimators, continuous refinement reduces the 30 dB NMSE
from $1.890\times10^{-3}$ for Grid NF-SPR to $7.219\times10^{-4}$,
a reduction of 62\%. Since both use the same channel family, this comparison
isolates the benefit of fitting continuous positions. The remaining gap to
the known-position genie reflects the cost of estimating geometric support.
Thus, the high-SNR gain comes from resolving and refining geometry under a
valid model, not from recovering an arbitrary vector in the null space of
$\vc\Phi$.

The right panel tests position coverage while keeping propagation unchanged.
At a 0.4 m shift, OG-NF-SPR attains $5.918\times10^{-3}$ NMSE, compared with
0.0312, 0.127, 0.0257, and 0.0247 for LMMSE, KNN, Compact VOR-DNN, and
DenseNet-VOR, respectively. Grid NF-SPR and the genie also retain low errors.
The physical atom map remains valid over the search region, while the
learned maps and stored statistics reflect the sampled position distribution.
Hence, source motion need not require retraining when the geometry remains
valid and the search region covers the new positions. This is a benefit of
position generalization within one deployment, not evidence of robustness
to a different propagation environment.

\subsection{Shared-Snapshot Estimates and Refinement Mechanism}

Fig.~\ref{fig:surfaces_paper} provides a signal-level view of the matched-model
comparison using three shared snapshots at 30 dB. Broad ridges in the coarse
score indicate poorly distinguished candidate positions; the refined MAP
profiles show the local search between grid points. The bottom row then
compares the reconstructed aperture magnitudes with the true spatial
variation. These panels connect the search procedure to the quantity the
algorithm is intended to recover.

Across these snapshots, OG-NF-SPR achieves complex NMSEs of
$(2.8,6.9,7.1)\times10^{-4}$, compared with $5.2\times10^{-3}$ to
$2.2\times10^{-2}$ for the deep reconstructors. The corresponding OG-NF-SPR
position errors are 0.051, 0.144, and 0.270 m. Small aperture error therefore
does not imply equally precise position recovery: weakly distinguishable
positions can produce similar reconstructed fields. This is why the
RIS-domain and position-domain metrics are reported separately.

The heat maps hold altitude at its estimated value, so they are slices of a
three-dimensional search, not full position-error maps. Likewise, magnitude
traces do not display phase errors; the reported complex NMSE does. These
distinctions prevent visual agreement in one panel from being mistaken for
accurate recovery of every channel component or position coordinate.

\subsection{Error Distribution and Calibration Sensitivity}\label{subsec:mismatch}

\begin{figure*}[t]
\centering
\includegraphics[width=0.99\textwidth]{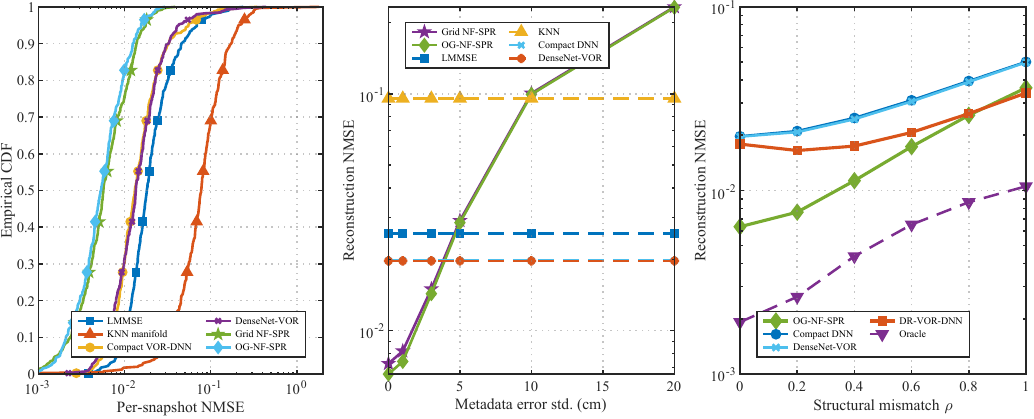}
\caption{Robustness at 20 dB. Left: matched-channel error CDF. Middle:
reflector-metadata error with an unchanged test channel. Right: structural
mismatch from \eqref{eq:structural_mismatch_paper}; at $\rho=1$, four omitted
paths and 4\%/$6^\circ$ RMS element gain/phase errors are active. DR-VOR-DNN
uses the Compact architecture and label budget but is trained over
$\rho\sim\mathcal U[0,1]$.}
\label{fig:robustness_paper}
\end{figure*}

The SNR sweep assumes that the physical model is correct.
Fig.~\ref{fig:robustness_paper} first checks how its gain is distributed across
snapshots and then relaxes that assumption in two ways: erroneous reflector
coordinates with an unchanged signal, and a changed physical channel.
The middle and right panels must therefore be interpreted separately.

The matched-channel CDF shows that the improvement is not limited to the
aggregate NMSE. OG-NF-SPR has a median per-snapshot NMSE of
$5.327\times10^{-3}$, compared with $1.320\times10^{-2}$,
$1.349\times10^{-2}$, and $1.812\times10^{-2}$ for Compact VOR-DNN,
DenseNet-VOR, and LMMSE. The gain is therefore also present near the center
of the error distribution.

The middle panel perturbs only the reflector coordinates supplied to the
physical estimators; the true channel and noisy BS samples remain unchanged.
OG-NF-SPR's NMSE increases from $6.553\times10^{-3}$ with exact coordinates
to $1.432\times10^{-2}$ at 3 cm and $2.857\times10^{-2}$ at 5 cm, crossing
the deep baselines between these points. Grid NF-SPR degrades similarly.
Incorrect reflected atoms affect both the initial projection and the final
channel fit, so continuous refinement of the UAV position cannot compensate
for arbitrary reflector-calibration errors.

In contrast, LMMSE, KNN, and the deep networks remain unchanged because neither
their BS inputs nor their stored statistics, labels, or weights are modified.
Their horizontal curves demonstrate independence from this metadata input,
not adaptation to unseen propagation. Moreover, the sweep scales one fixed
random realization of coordinate error. The crossing is a sensitivity result
for this deployment and error direction, not a universal 3--5 cm calibration
threshold.

\subsection{Structural Mismatch and Training Coverage}
\label{subsec:structural_mismatch}

The right panel of Fig.~\ref{fig:robustness_paper} asks a stronger question:
can learning help when the signal itself lies outside the nominal model?
The test channel now follows \eqref{eq:structural_mismatch_paper}. At
$\rho=1$, four omitted weak paths and 4\%/$6^\circ$ RMS element gain/phase
errors are present. The nominal physical and deep estimators are unchanged.
DR-VOR-DNN has seen this mismatch family during training, with the same
architecture and label budget as Compact VOR-DNN.

OG-NF-SPR retains lower NMSE than the learned estimators through $\rho=0.8$,
where its error is $2.568\times10^{-2}$ versus $2.623\times10^{-2}$ for
DR-VOR-DNN and approximately $3.9\times10^{-2}$ for the nominal networks.
At $\rho=1$, DR-VOR-DNN attains $3.397\times10^{-2}$, modestly improving
on OG-NF-SPR's $3.622\times10^{-2}$, while the nominal networks approach
$5.0\times10^{-2}$. The ``Oracle'' curve additionally knows the true
position, all reflected paths, and element responses, but still estimates
their gains. Its lower error indicates the remaining benefit of information
unavailable to the practical estimators.

The important comparison is between Compact VOR-DNN and DR-VOR-DNN:
architecture, position-sample count, and noise augmentation are unchanged,
yet including the discrepancy in training reduces the error. This ties the
gain to coverage rather than network size. It also explains why the nominal
networks do not automatically benefit when the physical model is inaccurate.
The modest ordering reversal at $\rho=1$ is specific to the tested
repeatable mismatch family, not a guarantee for unrelated environments.
The practical implication is to improve the relevance of the labels, rather
than assume that a larger network will compensate for missing channel
conditions.

\begin{figure*}[t]
\centering
\includegraphics[width=0.88\textwidth]{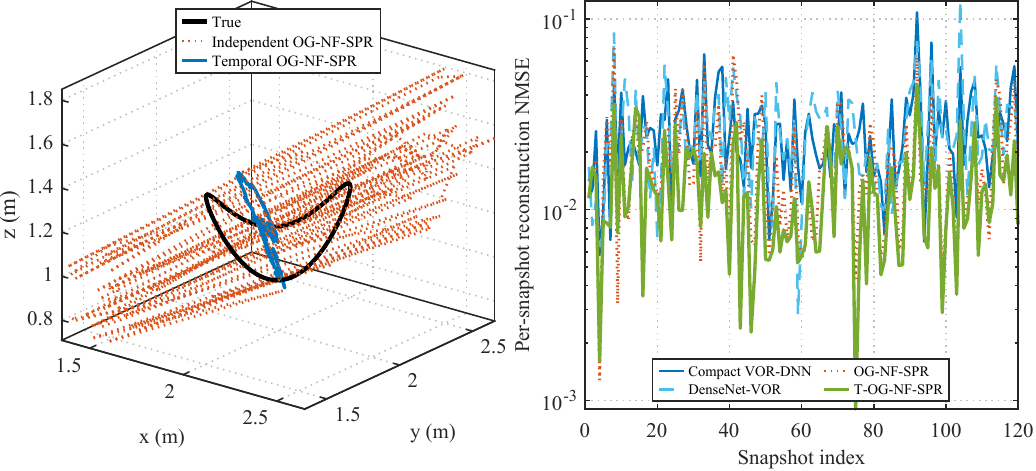}
\caption{Learning/modeling comparison on one shared trajectory at 15 dB.
Left: true trajectory and the positions estimated by independent OG-NF-SPR and
causal T-OG-NF-SPR with $\sigma_p=0.08$ m. Right: per-snapshot RIS-domain NMSE
of the frozen Compact VOR-DNN, frozen DenseNet-VOR, OG-NF-SPR, and T-OG-NF-SPR
under identical observations and noise.}
\label{fig:trajectory_paper}
\end{figure*}

\begin{figure*}[t]
\centering
\includegraphics[width=0.98\textwidth]{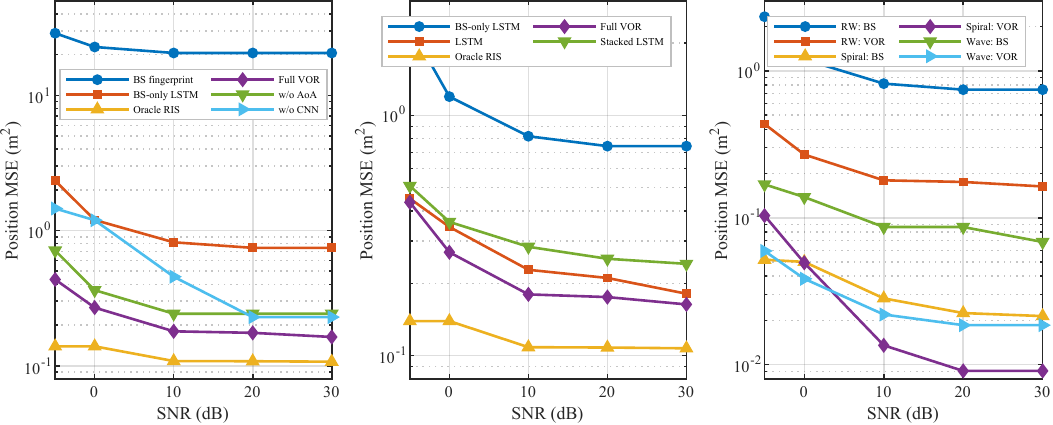}
\caption{Downstream tracking ablations of the VOR--SSFE--BiTP chain,
redrawn from the results in \cite{WuJSTSP2026}. Left:
complete VOR+SSFE+BiTP, BS-side references, oracle RIS, and feature ablations.
Middle: BiTP versus unidirectional LSTM variants. Right: VOR-assisted
and BS-only tracking for random-walk, spiral, and wave trajectories.}
\label{fig:deep_tracking_paper}
\end{figure*}

\subsection{Reconstruction Along a Shared Trajectory}

Fig.~\ref{fig:trajectory_paper} evaluates temporal information on a common
sequence at 15 dB. Compact VOR-DNN, DenseNet-VOR, independent OG-NF-SPR, and
causal T-OG-NF-SPR achieve trajectory NMSEs of
$(2.649,2.706,1.911,1.332)\times10^{-2}$, respectively. With
$\sigma_p=0.08$ m, temporal regularization also reduces position RMSE from
0.653 m to 0.225 m. Since the two physical estimators share the propagation
model and measurements, their difference isolates the use of the previous
position estimate. The gain is consistent with suppressing jumps between
ambiguous candidate locations, rather than increasing the number of
measurements in the current snapshot.

Only the physical estimators appear in the position panel because the deep
reconstructors output VORs, not positions. All four methods are compared in
the reconstruction panel using identical observations and noise. A tracker
such as \eqref{eq:deep_tracking_paper} is needed to convert learned VORs into
positions; its contribution should not be attributed to the reconstructor.

T-OG-NF-SPR uses the previous estimate and the current observation. A small
motion covariance strengthens temporal regularization but can resist a true
maneuver or propagate an earlier error, whereas a large covariance approaches
independent-snapshot fitting. The reported gain therefore establishes the
usefulness of the selected motion scale on this sequence, not superiority
under arbitrary dynamics. More elaborate motion priors require separate
validation.

\begin{table*}[t]
\centering
\caption{Prior Reliability and Deployment Choice}
\label{tab:comparison_paper}
\footnotesize
\begin{tabularx}{0.98\textwidth}{p{0.22\textwidth}|p{0.23\textwidth}|X}
\toprule
Operating condition & Supported choice & Evidence and interpretation \\
\midrule
Severe noise, covered distribution & LMMSE or Compact VOR-DNN & At $-5$ dB, NMSE is 0.450/0.494 versus 0.926 for OG-NF-SPR; averaging reduces support-selection errors. \\
Reliable geometry, informative observations & OG-NF-SPR & At 20 dB, NMSE is $6.553\times10^{-3}$ versus about $1.97\times10^{-2}$ for both networks; refinement exploits continuous geometry. \\
Position shift or unavailable labels & OG-NF-SPR & The physical atom map remains valid inside the search region. At 0.4 m shift, NMSE is $5.918\times10^{-3}$ versus 0.0247 for DenseNet-VOR. \\
Incorrect reflector metadata, unchanged channel & Compact or DenseNet-VOR & Networks do not consume the erroneous coordinates. The 3--5 cm crossover is specific to the tested deployment. \\
Strong repeatable mismatch covered by labels & DR-VOR-DNN & At $\rho=1$, NMSE is 0.0340 versus 0.0362 for OG-NF-SPR; nominal networks remain near 0.05. \\
\bottomrule
\end{tabularx}
\end{table*}

\subsection{Implications for Downstream Tracking}

Reconstruction is an intermediate step; its downstream value depends on the
tracker. Fig.~\ref{fig:deep_tracking_paper} revisits the tracking results of
\cite{WuJSTSP2026}, redrawn here to explain this distinction. Its separate
protocol uses 32 RIS profiles, 3000 trajectories per SNR/type, and a ten-slot
window. These are not new shared-data comparisons between OG-NF-SPR and the
deep reconstructors.

In the left panel, ``Full VOR'' denotes the complete VOR+SSFE+BiTP chain.
At 30 dB, its position MSE is $0.163$ m$^2$, compared with $0.243$ m$^2$
for ``w/o AoA'' and $0.229$ m$^2$ for ``w/o CNN''. Angular cues and spatial
feature extraction thus make complementary contributions in this setup.
The higher errors of ``BS fingerprint'' and ``BS-only LSTM'' show the
limitations of the tested direct BS-side representations. The ``Oracle RIS''
reference attains $0.107$ m$^2$ with direct RIS-domain input: removing
reconstruction error helps, but does not remove feature-extraction and
prediction error.

The middle panel isolates the predictor architecture. BiTP achieves
$0.163$ m$^2$ at 30 dB, versus $0.181$ and $0.241$ m$^2$ for the one-layer
and stacked unidirectional LSTMs. Thus, a deeper temporal model is not
necessarily better in this experiment. Bidirectional processing uses both
directions within the available window; since that window ends at the current
slot in \eqref{eq:deep_tracking_paper}, it does not require future samples.
These accuracy curves alone do not quantify inference latency.

The right panel shows that the value of VOR also depends on motion.
At 30 dB, VOR-assisted tracking reduces random-walk MSE from 0.745 to
$0.163$ m$^2$ and wave-trajectory MSE from 0.0684 to $0.0186$ m$^2$.
For a smooth spiral at $-5$ dB, however, BS-only LSTM achieves
$0.0518$ m$^2$, outperforming the full chain's $0.103$ m$^2$.
A plausible explanation is that a predictable trajectory is already well
captured by temporal processing, while noisy reconstruction introduces
additional error. These ablations motivate choosing reconstruction and
prediction separately. Establishing which reconstructor gives better
end-to-end tracking would require the same tracker protocol on a common
sequence dataset, including training for each reconstruction-error
distribution.

\subsection{Deployment Implications}

Table~\ref{tab:comparison_paper} summarizes three design implications.
First, SNR alone is insufficient: informative measurements favor physical
fitting only when its geometry is accurate. Second, position shift and
propagation change require different responses. A new position inside the
search region need not invalidate calibration, whereas changed reflectors
or omitted paths may require an updated model. Third, relevant training
coverage matters more than increasing label count or network depth without
covering the discrepancy.

A deployment should therefore assess calibration and labelled coverage,
then validate reconstruction and tracking separately under the expected
noise and motion. Reliable geometry enables OG-NF-SPR without paired
labels; representative labels can make mismatch-aware learning preferable
when the nominal model is deficient. If neither prior is reliable,
additional calibration or data are needed. The observed crossovers guide
this assessment but are not universal switching thresholds.

Learned initialization and residual correction are possible combinations,
not validated algorithms here. A learned seed cannot remove bias from an
incorrect final physical model, and residual learning still requires
representative labels.

\raggedbottom
\section{Conclusion}\label{sec:conclusion}

This paper studied how to reconstruct a virtual XL-RIS aperture signal
without direct surface measurements. We formulated learned inversion and
physical fitting under common BS observations and developed OG-NF-SPR as a
label-free alternative based on calibrated geometry. The numerical results
show that the preferred estimator depends on both measurement quality and
the validity of its prior. Statistical averaging is effective when noise
leaves several source locations plausible; continuous physical fitting is
effective when those locations can be distinguished under a valid model.
The mismatch experiments further separate the benefit of representative
training coverage from insensitivity to unused calibration metadata.
A causal extension demonstrates how temporal information can stabilize
reconstruction without trajectory labels. Together, these findings turn
the learning-versus-modeling comparison into a choice between explicit
deployment requirements, rather than an unconditional algorithm ranking.
Validation with measured channels and controlled combinations of the two
routes remain directions for future work.

\appendices
\section{Derivative of the Near-Field Atom}
From \eqref{eq:atom_paper},
\begin{equation}
\log[\vc a(\vc u)]_n=\log d_c-\log d_n-jk(d_n-d_c).
\end{equation}
Using
\begin{equation}
\frac{\partial d_n}{\partial u_i}=\frac{u_i-p_{n,i}}{d_n},\quad
\frac{\partial d_c}{\partial u_i}=\frac{u_i-p_{\mathrm R,i}}{d_c},
\end{equation}
and multiplying the logarithmic derivative by $[\vc a(\vc u)]_n$ yields
\eqref{eq:atom_derivative_paper}--\eqref{eq:log_derivative_paper}.

\bibliographystyle{IEEEtran}
\bibliography{myre}

\end{document}